\documentclass[a4paper,11pt]{article}
\pdfoutput=1 

\usepackage{jinstpub} 
\usepackage{graphicx}
\usepackage{subcaption}
\usepackage{multirow}
\usepackage{amsmath}
\newcommand{\bdec }{\mbox{$\beta$-decay}}

\title{\boldmath Technical design and commissioning of a sensor net for fine-meshed measuring of the magnetic field at the KATRIN spectrometer}

\author[a,b,1]{J. Letnev,\note{Corresponding authors}}
\author[a,d,2]{W. Hazenbiller, \note{previously W. Seller}}
\author[a,1]{A. Osipowicz,}
\author[c]{A. Beglarian,}
\author[c]{H. Bouquet,}
\author [d,e]{G. Drexlin,}
\author [d]{F. Gl\"uck}
\author[a]{J. Garbe,}
\author[b]{H. Hillmer,}
\author[a]{P. Marte,}
\author[d]{T. Th\"ummler,}
\author[f]{Ch. Weinheimer}

\affiliation[a]{University of Applied Sciences (HFD), Leipziger Str. 123, D-36037 Fulda, Germany}
\affiliation[b]{Institute of Nanostructure Technologies and Analytics (INA), University of Kassel, Heinrich-Plett-Str. 40, D-34132 Kassel, Germany}
\affiliation[c]{Institute for Data Processing and Electronics, Karlsruhe Institute of Technology, Hermann-von-Helmholtz-Platz 1, D-76344 Eggenstein-Leopoldshafen, Germany}
\affiliation[d]{Institute for Nuclear Physics (IKP), Karlsruhe Institute of Technology, Hermann-von-Helmholtz-Platz 1, D-76344 Eggenstein-Leopoldshafen, Germany}
\affiliation[e]{Experimental Particle Physics (ETP), Karlsruhe Institute of Technology, Hermann-von-Helmholtz-Platz 1, D-76344 Eggenstein-Leopoldshafen, Germany}
\affiliation[f]{Institut f\"ur Kernphysik, WWU M\"unster, Wilhelm-Klemm-Str. 9, D-48149 M\"unster, Germany}

\emailAdd{Johann.Letnev@et.hs-fulda.de}

\abstract{The KArlsruhe TRItium Neutrino experiment (KATRIN) aims to measure 
	the absolute neutrino mass scale with an unprecedented sensitivity of 0.2\,eV/c$^2$ (90\% C.L.), using $\beta$ decay electrons from tritium decay. The kinetic energy of the decay electrons is measured using an electrostatic integrating main spectrometer with magnetic adiabatic collimation and requires a certain magnetic field profile. For the control of the magnetic field in the main spectrometer area two networks of mobile magnetic field sensor units are developed and commissioned. The radial system is operated close to the outer surface of the main spectrometer whereas the vertical one 
	is mounted along vertical planes left and right of the main spectrometer. The sensor setup can take several thousand magnetic field samples at a fine meshed grid, thus allowing to study the  magnetic field inside the main spectrometer and the influence of 
	magnetic materials in the vicinity of the main spectrometer.}


\proceeding{}

\begin{document}
	\maketitle
	\flushbottom
	
	\section{Introduction}
	\label{sec:intro}
\begin{figure}[h]
		\centering 
		\includegraphics[width=.9\textwidth]{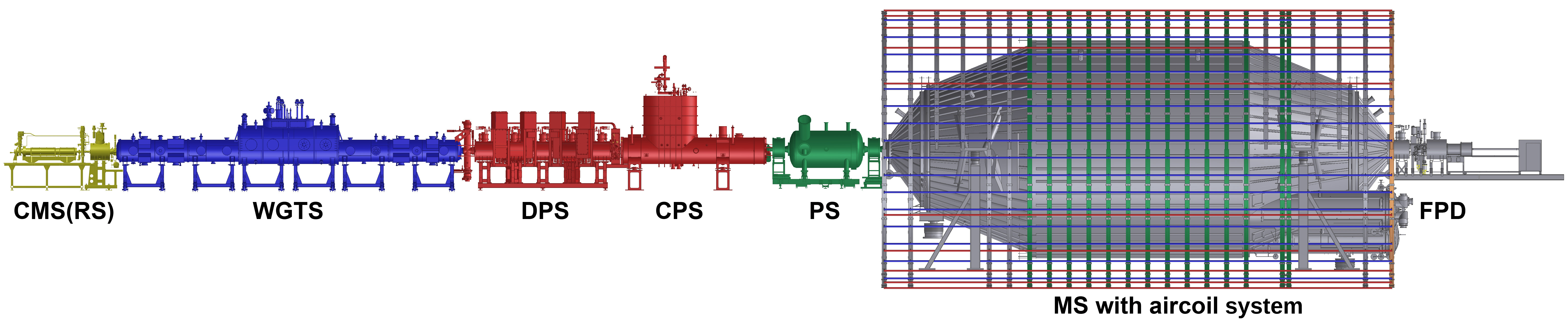}
		\caption{\label{fig:KATRINsetup} Schematic overview of 70m long KATRIN setup, consisting of calibration and monitor rear system (CMS/RS), the electron emitting section with the windowless gaseous tritium source (WGTS), differential pumping (DPS) and cryotrapping section (CPS), the small pre-spectrometer (PS) and the large main spectrometer (MS) with the aircoil system and lastly the segmented PIN-diode detector with upstream pinch and detector magnet (FPD).}
	\end{figure}

\begin{figure}[h]
		\centering 
		\includegraphics[width=.75\textwidth]{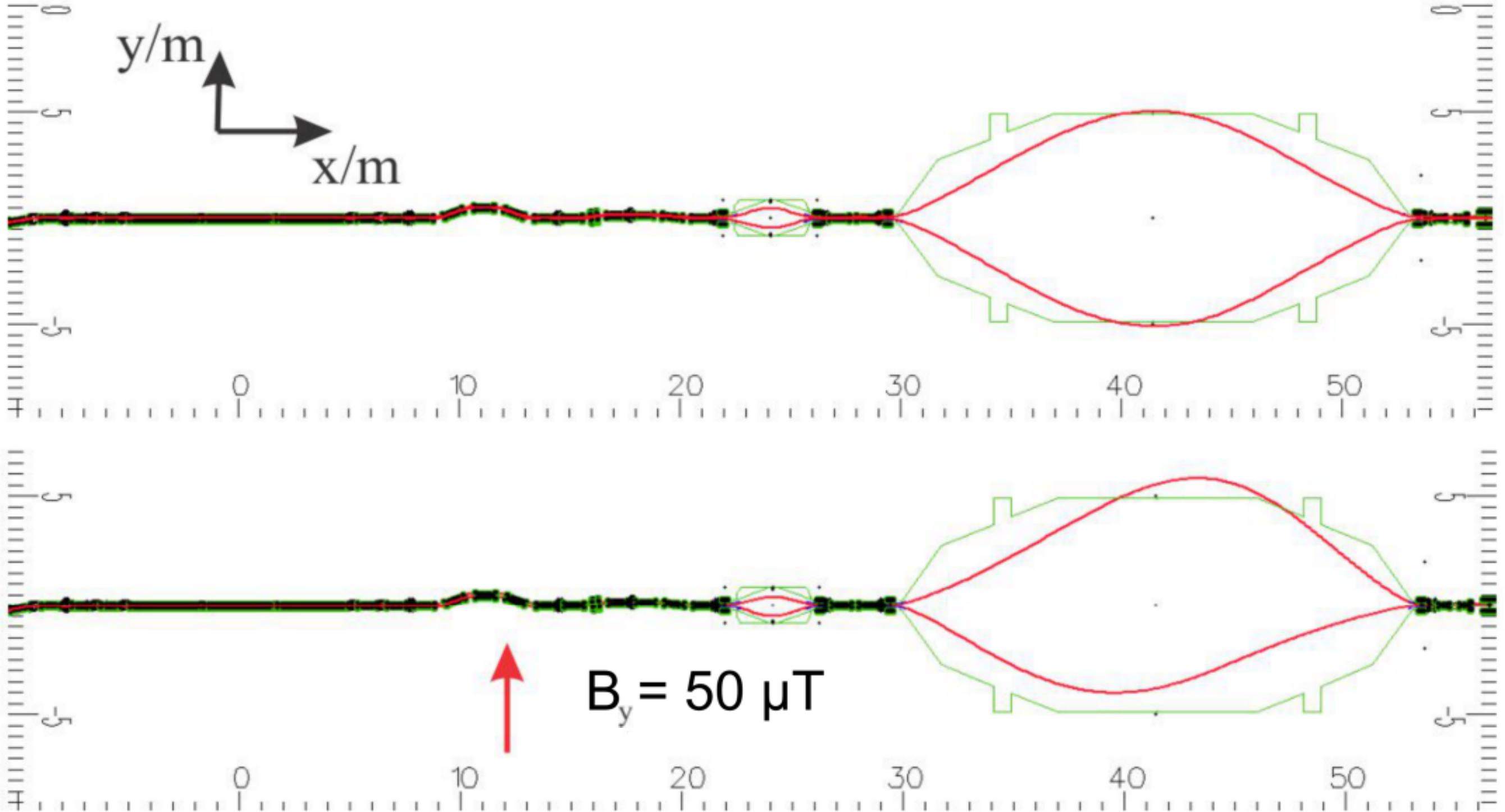}
		\caption{\label{fig:MagLinesKAT}A CAD view (taken from a PartOpt \cite{Zip2002} simulation)  showing  the KATRIN high field solenoids energized according to reference values, and in red the corresponding boundary $190$ Tcm$^2$ magnetic transport flux lines in the central horizontal plane without air coils. Both axes are given in meters. Top: without external perturbation touching the spectrometer walls. Bottom: In case of an additional global external magnetic field (e.g.earth magnetic field) of $By = 50$ $\mu$T the perturbed field lines intersect with the MS walls.}
	\end{figure}

\begin{figure}[h]
		\centering 
		\begin{subfigure}[b]{0.55\textwidth}
			\includegraphics[width=\textwidth]{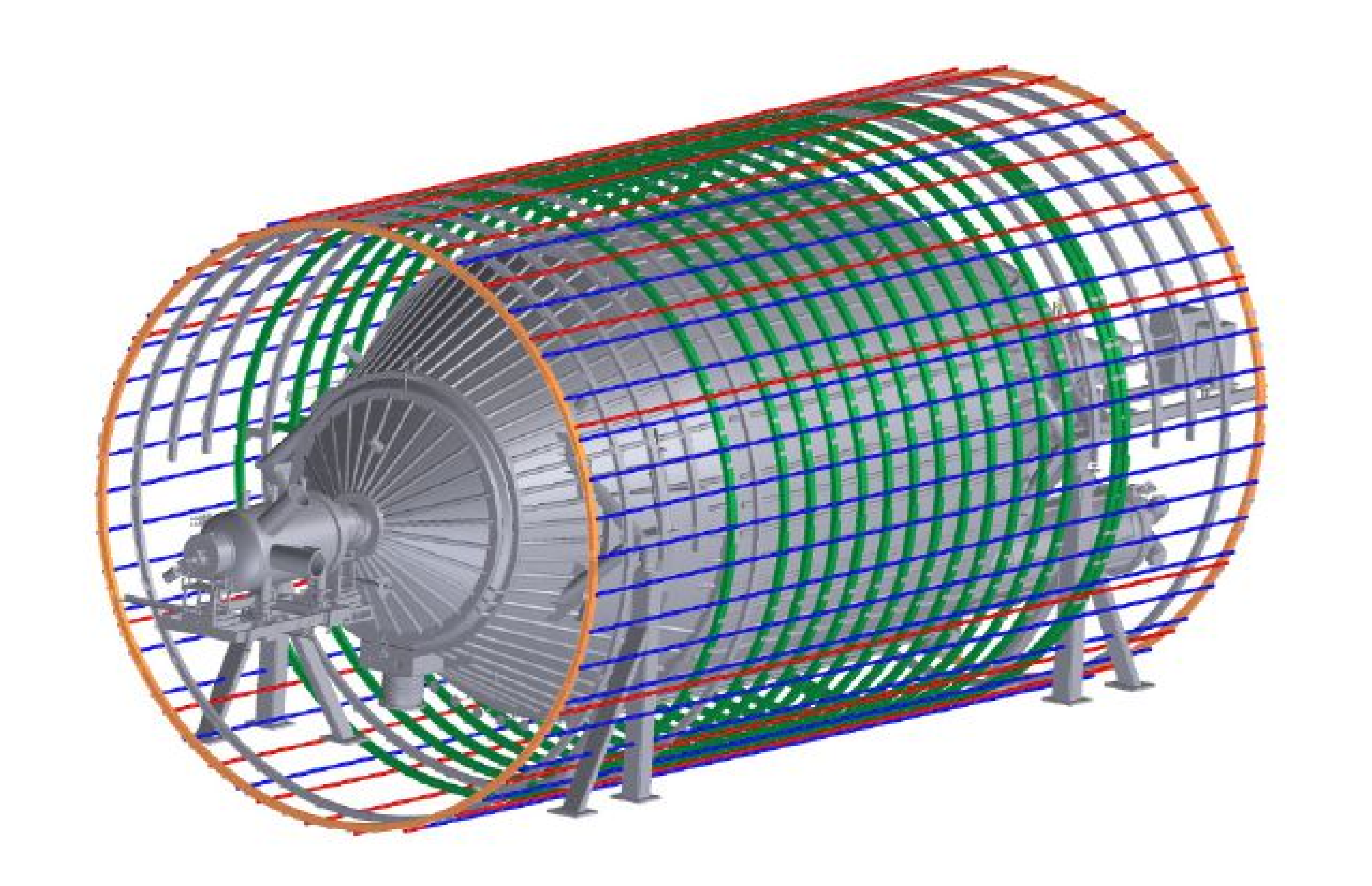}
			\caption{}
			\label{fig:gull1}
		\end{subfigure}
		\quad
		\begin{subfigure}[b]{0.38\textwidth}
			\includegraphics[width=\textwidth]{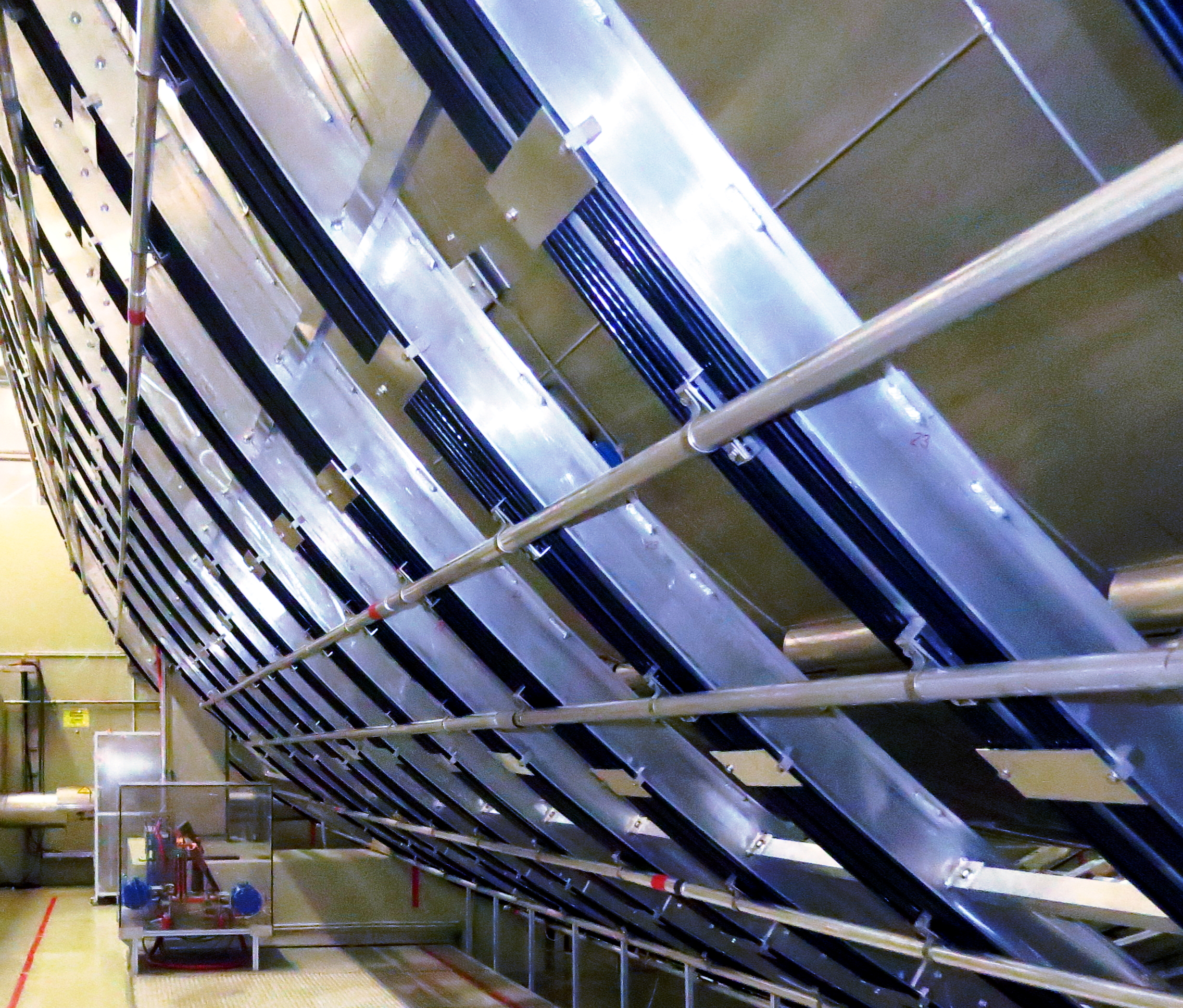}
			\caption{}
			\label{fig:gull2}
		\end{subfigure}
		\caption{\label{fig:Aircoilsystem} The CAD (a) and photographic (b) view of KATRIN main spectrometer. LFCS: The ring shaped low field coils distributed co-axially along the MS (green circles in Figure (a)). The current leading elements of the EMCS run parallel to the spectrometer walls (blue and red lines as well as orange circles in Figure (a)). A more detailed description can be found in \cite{Erh2018}.}
	\end{figure}

	The Karlsruhe TRItium Neutrino experiment \cite{FZK2004} is a next-generation experiment for a direct and model-independent determination of the absolute neutrino mass scale
	. By analyzing the shape of the tritium \bdec~spectrum  near the endpoint energy at $E_0 = 18.6$ 
	KATRIN  will achieve a sensitivity of $m_{v}= 0.2 $ eV/c$^2$ (90\% C.L.). A schematic overview of the KATRIN setup is shown in \autoref{fig:KATRINsetup}.
	The experimental setup uses a magnetic transport flux of 190\,Tcm$^2$ to guide the \bdec~electrons from a windowless gaseous tritium source through a pumping section towards two electrostatic energy spectrometers and onto a detector. The operating principle of the spectrometers is based on a magnetic adiabatic collimation with electrostatic filtering (MAC-E filter)\cite{Lob1985, Bac1987, Pic1992}, where a retarding electric potential is used to reflect electrons below a given energy threshold. In order to ensure the correct function of the MAC-E filter, a certain magnetic field profile is required. The shape of the magnetic flux tube inside the main spectrometer (MS) 
	has a significant influence on the overall energy resolution function of the spectrometer. In addition, the alignment and shape of the magnetic field lines plays an essential role for the electronic background via a) the generation of secondary electrons through wall contact of energetic electrons (see \autoref{fig:MagLinesKAT}) and b) the generation and storage of charged particles due to penning traps and the magnetic bottle effect.
	For the control of the desired magnetic field shape, large aircoil
	systems \cite{Glu2008, Erh2018} are arranged around the MS: The earth magnetic field compensation system (EMCS) for the compensation of the earth magnetic field and the low field coil system (LFCS) for the fine tuning of the magnetic transport flux tube (see \autoref{fig:Aircoilsystem}). The requirements on the magentic field and achieved performances of the magnetic field generating systems are described in more detail in \cite{Gil2018}. Although the calculation of the magnetic field inside the main spectrometer generated by all the relevant current leading elements is in principle possible and well performed, perturbing external dipoles, magnetization effects in the direct environment of the spectrometer and the incorrect alignment and orientation of the spectrometer solenoids, EMCS and LFCS can have a disturbing influence. Due to the extreme vacuum conditions the installation of magnetic field sensors inside the main spectrometer is not possible during KATRIN operation.
	
	This paper focuses on the technical realization of two magnetic sensor networks that allow to measure the magnetic field in the direct environment of KATRIN main spectrometer over large areas with fine meshed sample positions. The radial magnetic measuring system (RMMS), based on the mobile sensor unit \cite{Osi2012}, is operated on 4 LFCS rings. The vertical magnetic measuring system (VMMS) covers vertical planes parallel to the MS beam axis. 
	
	\section{The radial magnetic field measuring system}
\begin{figure}[h]
		\centering 
		\includegraphics[width=.6\textwidth]{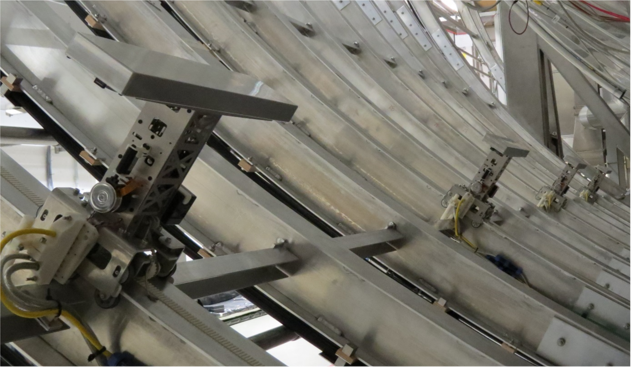}
		\caption{\label{fig:MobSUFoto} The four mounted mobile sensor units in their rest positions (docking stations) on the inner belts of the LFCS support structure. In the foreground the MobS unit on LFCS 3 close to the source side of the cylindrical part of the MS. Further away the MobS units on LFCS 6, 9 and 12 can be seen. }
	\end{figure}

\begin{figure}[h]
		\centering 
		\includegraphics[width=.85\textwidth]{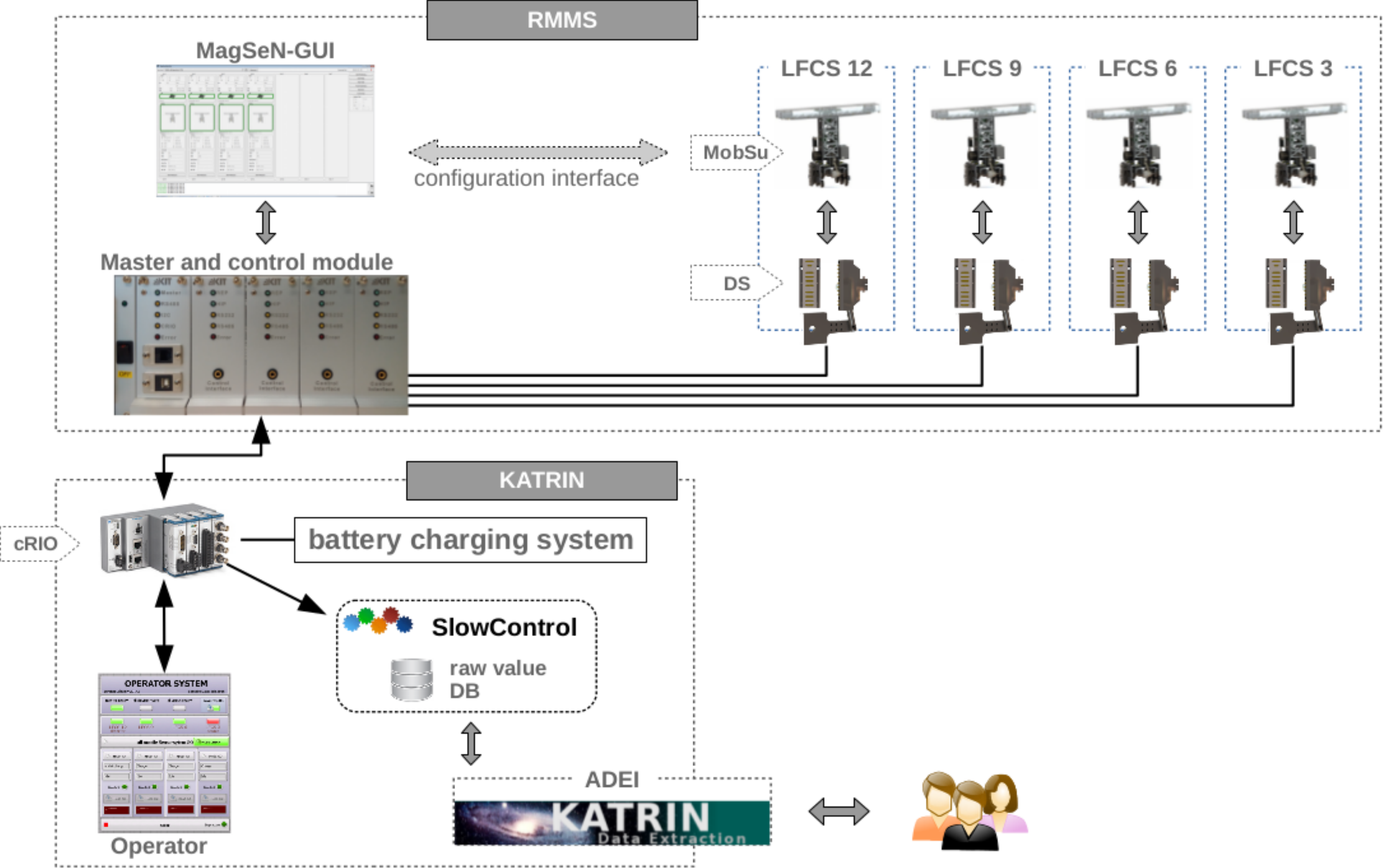}
		\caption{\label{fig:RMMSConfig} Schematic structure of the RMMS and its integration into the KATRIN slow control system. Figure is taken from \cite{Let2018}.}
	\end{figure}

	The radial magnetic field measuring system (RMMS) is a system for measuring the magnetic field close to the KATRIN MS surface. The initial concept is based on a mobile sensor unit (MobSU) \cite{Osi2012}, which moves on the inner side of the LFCS support ring and measures the magnetic field on predefined  sampling positions. According to the mechanical structure of the LFCS, up to 14 units can be installed. At present  four of these units have been installed and fully commissioned on LFCS 3, 6, 9 and 12 (see \autoref{fig:MobSUFoto}). This configuration has been chosen to get magnetic field values at  points symmetric with respect to the analyzing plane which is characterized by the minimal magnetic field $B_A$ at the center of the MS.
	\autoref{fig:RMMSConfig} displays the schematic interaction of all involved RMMS subsystems and their integration with the KATRIN experiment. The upper part of the figure shows the structure of radial magnetic measuring system with the master and control module and the four installed MobSU. 
	The so-called docking station (DS) is the start and end point of unit motion. It also represents the electromechanical as well as the data transfer link between the sensor unit and the master module (see \cite{Osi2012}) which is the interface to the KATRIN slow control database. Each subsystem of the RMMS can be configured and controlled by means of the PC tool 'MagSeN-GUI'. The connection between the radial magnetic field measuring system and the data management system and the SlowControl \cite{Beg2013} of KATRIN is realized via a modular CompactRIO Platform{\textsuperscript{\textregistered}}\footnote{CompactRIO is a registered trademark of National Instruments} (cRIO) \cite{NI}. In addition to the communication and transmission of the data to the higher-level processing stage, the controlled charging of the batteries installed on the MobSU is also performed via an interface integrated in the cRIO. Equivalent to many other KATRIN subsystems, the measured data of the RMMS can be accessed via an ADEI interface \cite{Beg2013}. 
	The complete system is described in \cite{Let2018}.
	
	\subsection{The Mobile Sensor Unit}
\begin{figure}[h]
		\centering 
		\includegraphics[width=.85\textwidth]{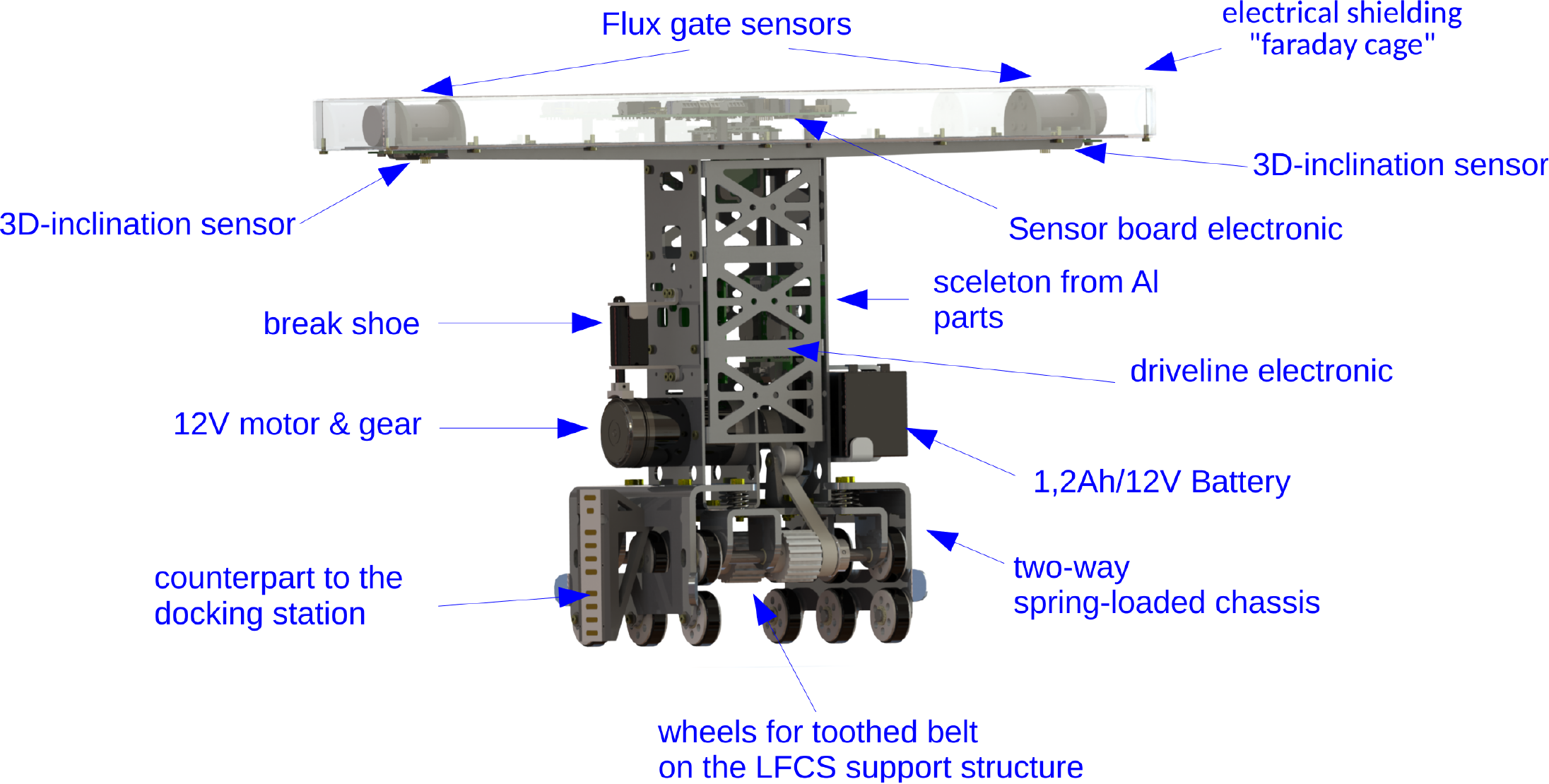}
		\caption{\label{fig:MobSU} Mechanical structure of the mobile sensor unit with the drive, frame and wings on which the flux-gate-sensors and the inclinometers are attached. Height: 296\,mm, Width: 532\,mm, Weight: 2.9\,kg. Distance between both flux-gate-sensors: 450\,mm}
	\end{figure}

\begin{figure}[h!]
		\centering
		\includegraphics[width=0.6\textwidth]{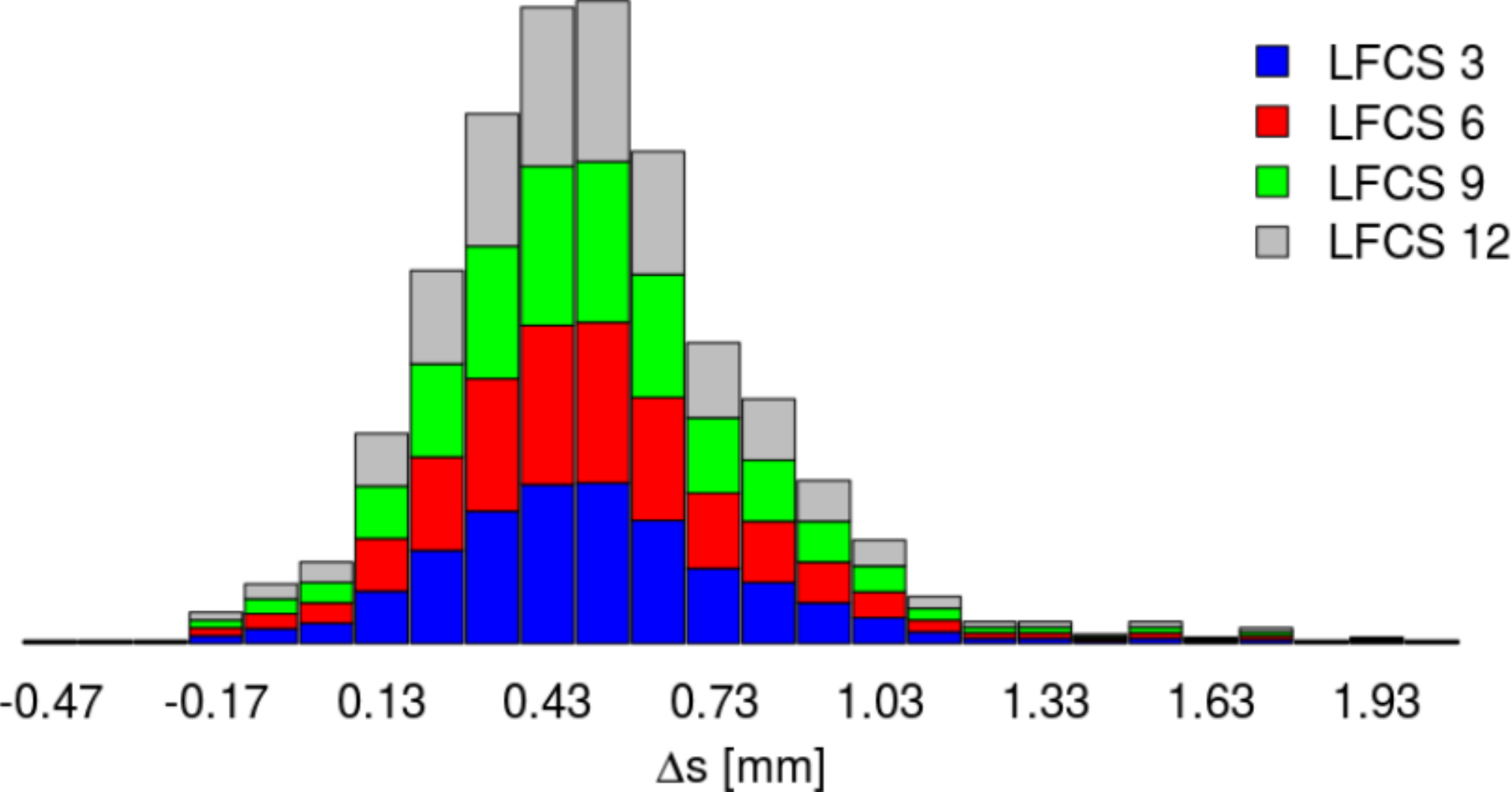}
		\caption{Distribution of the global stopping reproduction accuracy of the installed mobile units for 15 randomly selected measurement runs. Figure taken from \cite{Let2018}.}
		\label{fig:MobSUStopAccur}
	\end{figure}

	The mobile sensor unit represents the actual sensor from the point of view of the sensor network. The prototype of the unit described in \cite{Osi2012} has been modified and its properties improved. \autoref{fig:MobSU} shows the structure of the final MobSU version. The drive principle is now based on a combination of a tooth belt attached to the inner side of the LFCS support and toothed gear wheels within the MobSU drive. Due to the use of an aluminum skeleton layout of the drive chassis, the frame and the wings, a total weight of 2.9\,kg is achieved with a unit height of 296\,mm. The aluminum frame forms a Faraday cage and provides the necessary stiffness and electrical safety of the entire unit. The improved two-way spring-loaded chassis provides enough grip and dynamics to overcome the LFCS carrier's structural height, lateral offsets  and mechanical discontinuities along the track. 
	Furthermore, three-dimensional inclination sensor systems based on the FXLS8471Q 
	\cite{FXL} are positioned on the wings in such a way that they are centered parallel to the flux-gate-sensors\footnote{custom designed sensors FL3-1000 by Stefan Mayer Instruments with an accuracy of $\pm 0.5\%$ within the range of $\pm 1000\mu T$}. Due to a variation 
	in the temporal behavior of the individual components, especially due to a dependence on the battery voltage, 
	deviations of the positioning accuracy of the unit were detected. In order to counteract these uncertainties, a control algorithm without time dependent parameters has been developed, which is described in detail in \cite{Let2018}.
	By use of an incremental encoder, the local positions of the units on their tracks are recorded with a mechanical accuracy of 48.9\,$\mu$m and digital inclination accuracy of 0.37$^\circ$ during the entire revolution around the spectrometer. \autoref{tab:accurMobs} shows the determined sensor system accuracies.
	\begin{table}[h!]
		\centering 
		\begin{tabular}{|c|c|}
			\hline 
			\textbf{Sensor} & \textbf{Accuracy} \\ 
			\hline 
			\hline
			Magnetometer & 0.5\% (at $\pm 1000\mu$T) $\pm$20nT \\ 
			\hline 
			Position &  48.9$\mu$m $\pm$36nm\\ 
			\hline 
			Inclination &  0.37$^\circ$ $\pm$0.0219$^\circ$\\ 
			\hline 
		\end{tabular}
		\caption{Experimentally determined parameters for RMMS measurement accuracy after calibration process}
		\label{tab:accurMobs}
	\end{table}
	In addition, the maximum speed is reduced in a controlled manner for the area of 
	downward motion of the mobile unit. This procedure makes it possible to reach the target position with an accuracy better than 1\,mm. \autoref{fig:MobSUStopAccur} shows the 
	distribution of stopping 
	accuracy (difference between the target and reached position) for all four MobSU based on 15 randomly selected measurement runs.

\begin{figure}[h!]
		\centering 
		\centering 
		\begin{subfigure}[b]{0.4\textwidth}
			\begin{tabular}{|c|c|c|c|c|c|c|c|}
				\hline
				\textbf{number of pts.} & \textbf{meas. duration}\\
				\hline
				\hline
				60 & $\approx 9.5$\,min\\
				\hline 
				72 & $\approx 10.8$\,min\\
				\hline
				144 & $\approx 15.8$\,min\\
				\hline
			\end{tabular}
			\vspace{1.5cm}
			\caption{}
			\label{fig:MeasDur}
		\end{subfigure}
		\quad
		\begin{subfigure}[b]{0.5\textwidth}
			\includegraphics[width=\textwidth]{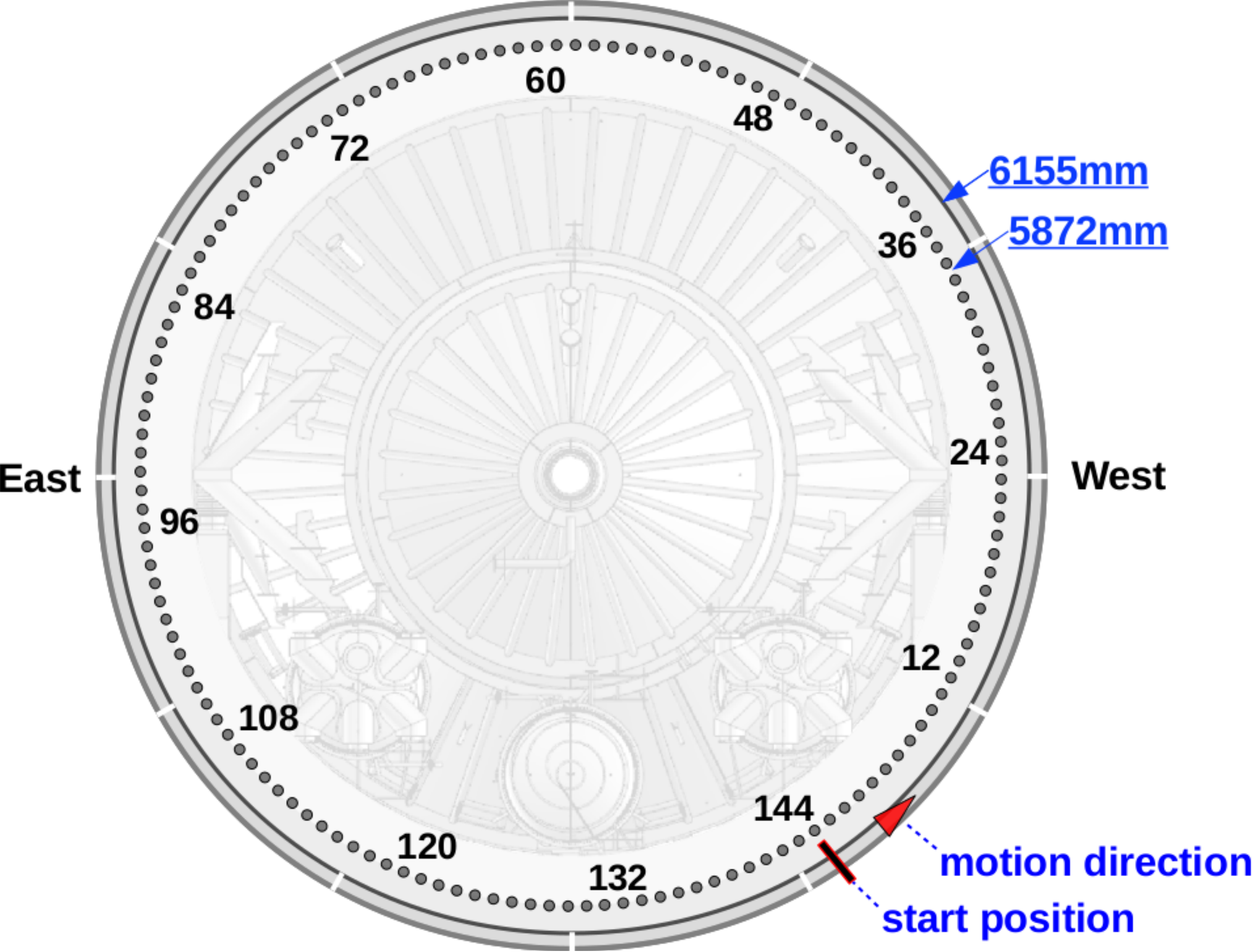}
			\caption{}
			\label{fig:MeasPosA}
		\end{subfigure}
		\caption{\label{fig:MeasPos} (a) Experimentally determined measurement duration for one measuring run depending on the configured number of positions. (b) Schematic representation of the sampling positions (dark circles) and sampling point numbers of the mobile sensor unit on the LFCS coil support structure with an average radius of 6155\,mm viewed from the detector site. The average radius of the flux-gate-sensor path is 5872\,mm. Starting position and motion direction are indicated.}
	\end{figure}

	In order to achieve the mechanical precision mentioned above, new concepts have been implemented in the motion control, as described in \cite{Let2018} in more detail. The necessary parameters are configured within the PC-Tool MagSeN-GUI. In particular, the desired number of magnetic field sampling positions on the entire LFCS ring must be set. The number of sampling positions (schematically shown in \autoref{fig:MeasPosA}) is used to calculate the required distance between the stopping positions of the mobile sensor unit which affects the total duration of one measurement cycle (see  \autoref{fig:MeasDur}). 
	
	\subsection{Magnetization effect and influence of compensation systems} 
\begin{figure}[h]
		\centering
		\begin{subfigure}[b]{0.46\textwidth}
			\includegraphics[width=1\textwidth, angle=0]{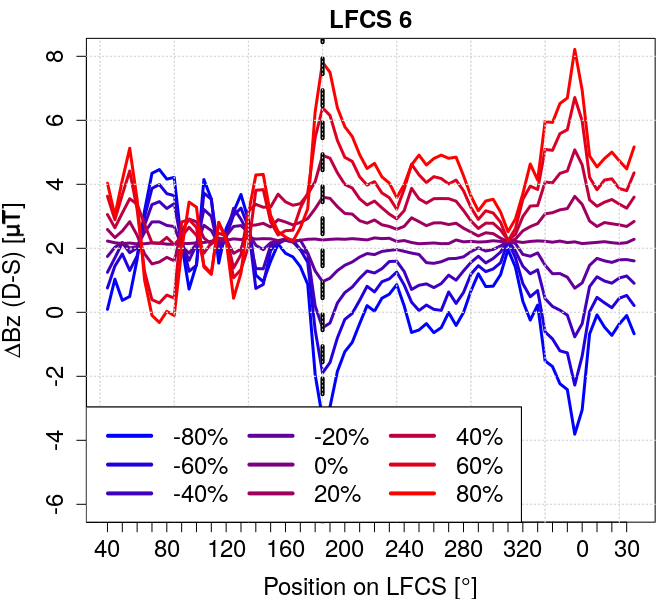}
			\caption{}
			\label{fig:hystdBz}
		\end{subfigure}
		\qquad
		\begin{subfigure}[b]{0.46\textwidth}
			\includegraphics[width=1\textwidth, angle=0]{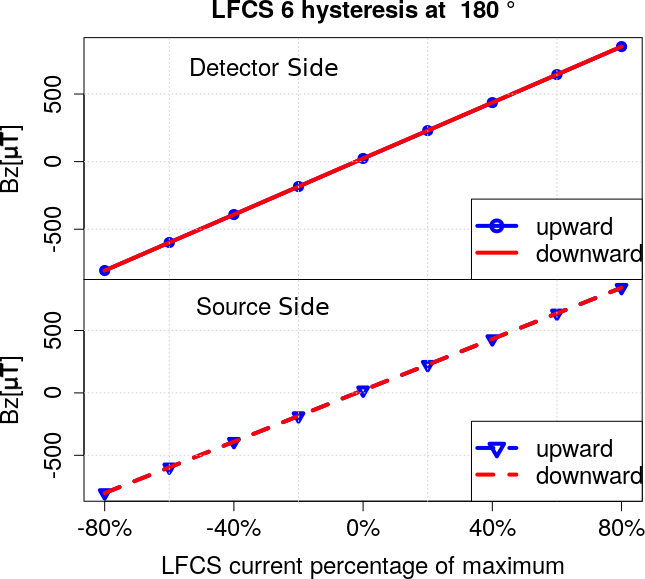}
			\caption{}
			\label{fig:hystdBzb}
		\end{subfigure}
		\caption{a) Absolute difference for $B_z$-component of the magnetic field between the two MobSU magnetometers for different percentages of  LFCS current settings in steps of $20\%$ of the full current. Full currents used (100\%): 100\,A for \textit{LFCS 1,2,12,13}, 115\,A for \textit{LFCS 3...11} and 70\,A for \textit{LFCS 14}. The MobSU position is shown in degree of rotation along the axis of the abscissas. Figure taken from \cite{Let2018}. b) Hysteresis view at position of 180$^\circ$ on the LFCS 6.}
		\label{fig:hystdBz1}
	\end{figure}

	During the commissioning phase of the large air coil system (see \cite{Erh2018}), the operational performance and functionality of the radial magnetic field sensor net was also inspected. For this purpose, the amperage of each individual LFCS coil 
	was gradually adjusted\footnote{in 20\% steps of the maximum permissible amperage (see \cite{Glu2008, Erh2018})} and the magnetic field 
	was recorded using the RMMS at $2*144$ measuring positions per MobSU. The 1152 points in total served as a basis for the investigation of possible magnetization effects. \autoref{fig:hystdBz} shows the absolute difference of the $B_z$-component\footnote{beam axis of the main spectrometer} of the magnetic field between the two flux gate magnetometers of a single sensor unit depending on the current in the associated LFCS coil, using  LFCS 6 as an example. The black dashed line indicates the position of the slice for the hysteresis view in \autoref{fig:hystdBzb}. It should be noted that all values used are in inclination corrected local MobSU coordinates.  

	\subsection{Coordinate system transformation and field determination in the analyzing plane}
\begin{figure}[h]
		\centering
		\includegraphics[width=.65\textwidth, angle=0]{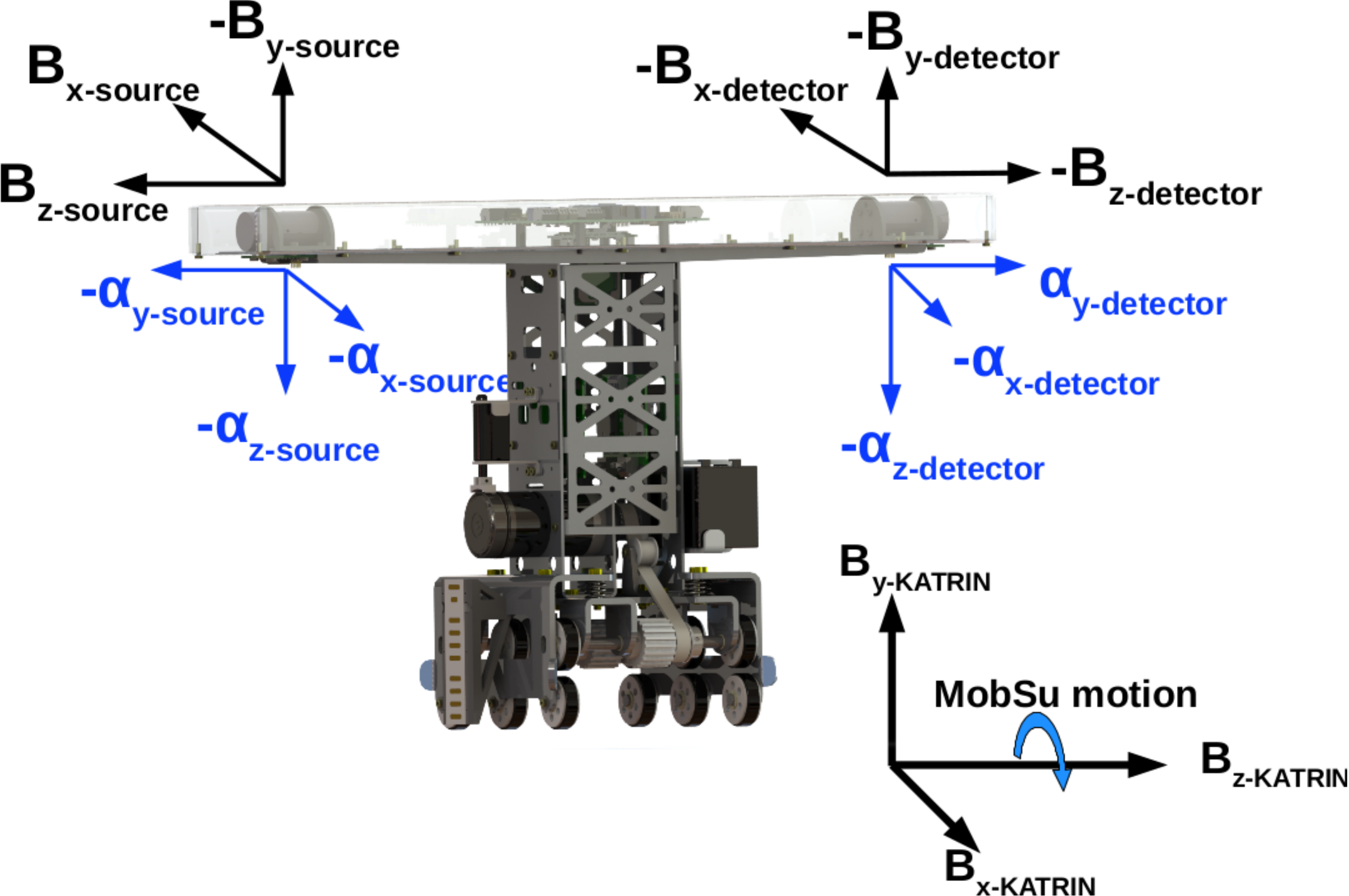}
		\caption{Orientation of the sensor components of the mobile sensor unit. Movement sequence of the unit around the spectrometer corresponds to the rotation around the longitudinal center axis of the MS (the z-axis of the global KATRIN coordinate system). Figure adopted from \cite{Let2018}.}
		\label{fig:mCoords}
	\end{figure}

\begin{figure}[th]
		\centering
		\includegraphics[width= .75\textwidth]{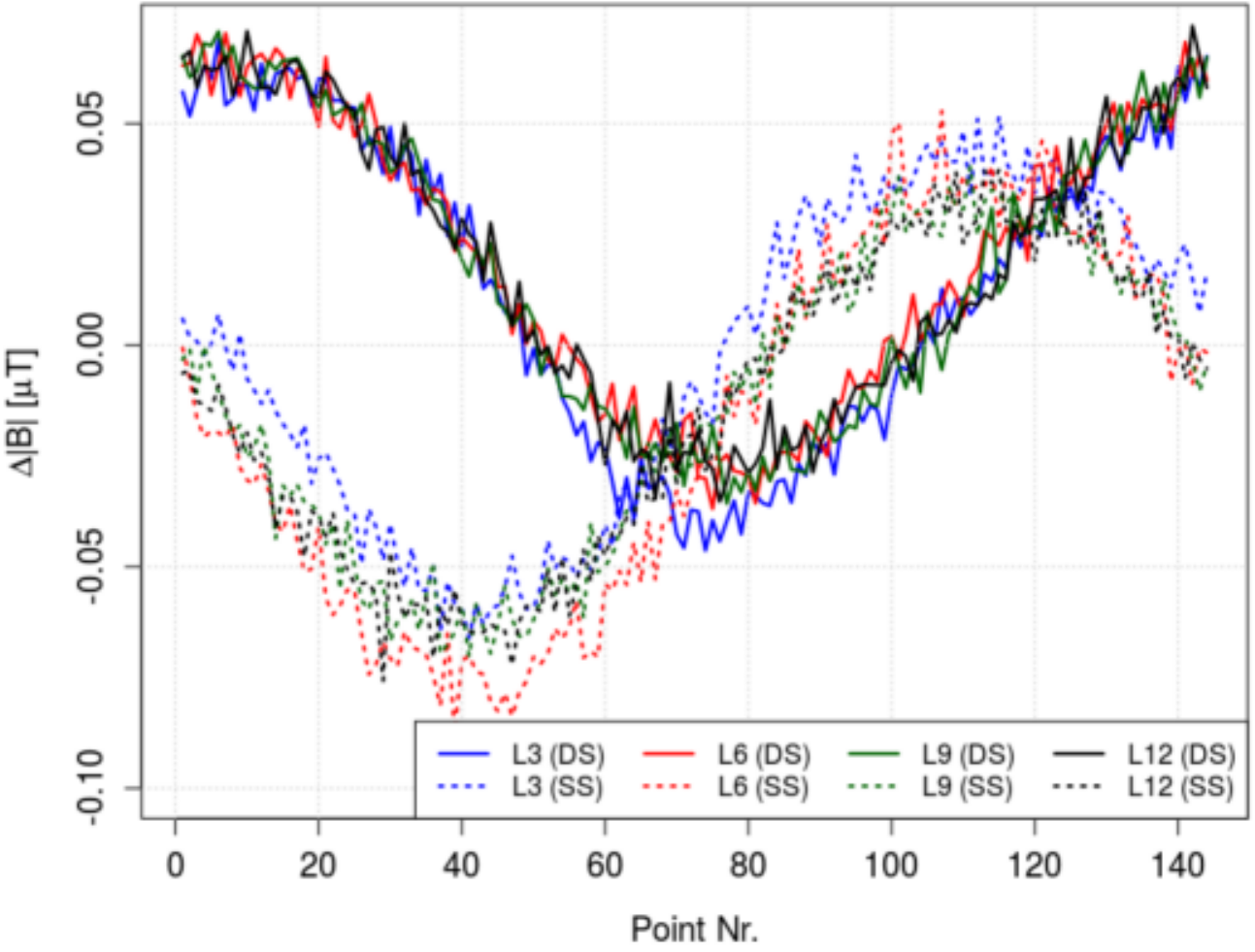}
		\caption{Total error of the magnetic field $\Delta |B|$ after transformation into the KATRIN coordinate system. Figure based on \cite{Let2018}.}
		\label{fig:BFehler}
	\end{figure}

	Due to the slight deformations  and a possible misalignment of the LFCS (see \cite{Erh2018}) the transformation of the locally obtained magnetic field values into a more global KATRIN coordinate system  \cite{Glu2009} is problematic. However, on the basis of the LFCS deformation measurement \cite{Bah2009} (data listing the radii of the LFCS at 36 angles along the circumference) a first attempt has been made.
	The deviation of the LFCS radii from the ideal $R_i= 6.155$ m at the sampling positions can be approximated iteratively by a spline interpolation taking into account the manually determined start positions and angles at the docking station.
	Based on this, the distance traveled by the sensor unit can be taken as the arc length $S$ of the LFCS circle to determine the global position $t_{x,y,z}$ of the unit under the condition $S_{Model} \equiv S_{MobSU}$.
	\begin{equation}
	S_{Model}= A_{corr} \cdot
	\int_{\alpha_{Dock}}^{\alpha_{th}}
	\sqrt{
		\left(R^2 + \left({{\partial R} \over {\partial\alpha}}\right)^2
		\right)_{\alpha = u}
	}  du
	\label{eq:MathMod}
	\end{equation}
	\begin{equation}
	t_x = R\cdot cos(\alpha_{th}) \qquad t_y = R\cdot sin(\alpha_{th}) \qquad t_z = t_z
	\label{eq:MobPosEq}
	\end{equation}
	$A_{corr}=\frac{s_{Dock}} {s_{Model}}$ represents a standardization constant at $\alpha_{th} = \alpha_{Dock}$ condition where $\alpha_{th}$ indicates the numerically determined theoretical rotation angle of MobSU. The experimentally determined positional data for the position in the z-direction $t_z$, the LFCS total circumference\footnote{total travel distance of the mobile sensor unit} $S_{Dock}$ and the start or end angle $\alpha_{Dock}$ of the MS revolution are summarized in \autoref{tab:PosDataForSpline}. 
	\begin{table}[t]
		\centering
		\begin{tabular}{|l|c|c|c|c|}
			\hline 
			& LFCS 3 & LFCS 6  & LFCS 9 & LFCS 12 \\ 
			\hline \hline
			\textbf{$t_z$} [m] ([mm]) & -4.040 (5) & -1.338 (5) & 1.354 (5) & 4.058 (5) \\ 
			\hline 
			\textbf{$s_{Dock}$} [m] ([mm])& 38.715 (3.96)  & 38.705 (3.89)  & 38.745 (3.11)  & 38.678 (3.23)  \\ 
			\hline 
			\textbf{$\alpha_{Dock}$} [$^\circ$]& 36.86 (0.175) & 36.04 (0.216)  & 37.03 (0.307) & 37.07 (0.349) \\ 
			\hline 
		\end{tabular} 
		\caption{Data of the LFCS support beams : $t_z$ the position in z direction relative to the MS center,   $S_{Dock}$ the total arc of sensor path for 1 revolution and $\alpha_{Dock}$ the start value for the inclination. The numbers in brackets represent the absolute deviation determined by using several unit runs.}
		\label{tab:PosDataForSpline}
	\end{table}
	With the sensor element orientation shown in \autoref{fig:mCoords}, the known theoretical rotation angle $\alpha_{th}$ and the measured inclination angles $\vec{g}_{DS}$ and $\vec{g}_{SS}$\footnote{DS for detector sided sensor and SS for source sided sensor of the mobile sensor unit} of both MobSU magnetometer, the inclination corrected rotation matrix $M_{rot}$ can be created.
	\begin{equation}
	M_{rot} =  N_{MobSU} \cdot N_{Trans} \cdot K
	\end{equation}
	Where $N_{MobSU}$ represents an ideal rotation matrix based on $\alpha_{th}$, $N_{Trans}$ specifies the MobSU construction-related translation matrix\footnote{displacement and orientation in relation to the MobSu magnetometers} and $K$ indicates the inclination correction matrix. A more detailed description is given in \cite{Let2018}. The coordinate transformation has a considerable influence on the total error of the magnetic field measurement, $\overrightarrow{\Delta B}$ which is calculated according to \autoref{eq:BError} and is shown in \autoref{fig:BFehler}. It should be noted that all of the magnetic field errors on the Detector Side (DS) and the Source Side (SS) correlate in different ways. This can be explained by inaccuracies and deviations of the mathematical model of coordinate transformation presented here. As the overall error is relatively small, this aspect can be neglected. To achieve better results, an improvement of the model data from \cite{Bah2009} by at least a factor of 10 is necessary.
	\begin{equation}
	\overrightarrow{\Delta B} = M_{rot_{DS/SS}} \cdot {{\overrightarrow{\sigma_B}_{DS/SS}} \over {\sqrt{32}}} + \overrightarrow{\Delta g}_{max} \cdot |B_{DS/SS}|
	\label{eq:BError}
	\end{equation}
	
	$M_{rot}$ specifies the mentioned corrected rotation matrix for the individual magnetometer, $\overrightarrow{\sigma_B}$ represents the corresponding uncertainty 
	of the magnetic field measurement and $\overrightarrow{\Delta g}_{max}$ indicates the maximum error of the MobSU internal inclination system.
	
\begin{figure}[h!]
		\centering
		\begin{subfigure}[b]{0.45\textwidth}
			\includegraphics[width=1\textwidth]{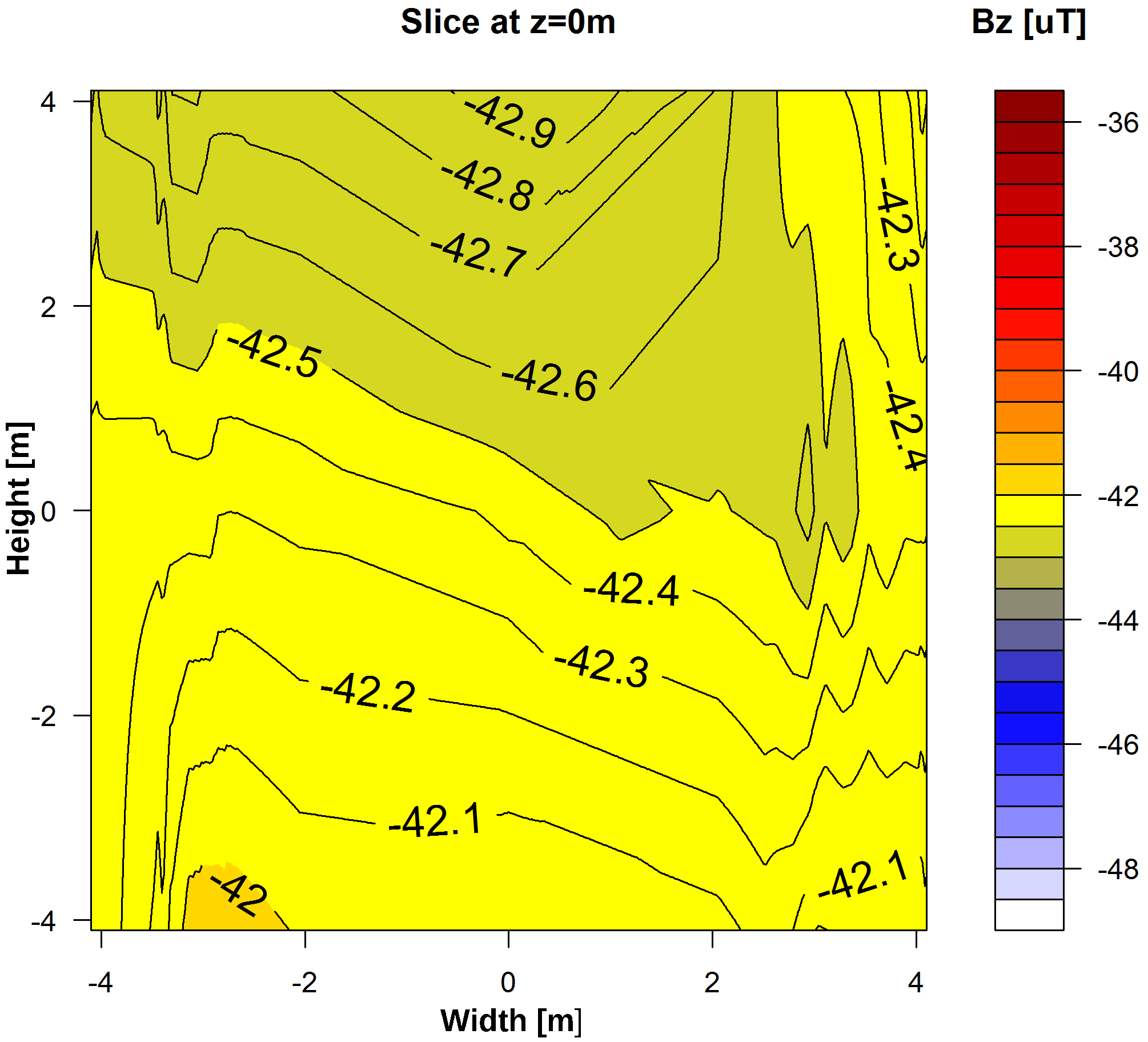}
			\caption{}
		\end{subfigure}
		\quad
		\begin{subfigure}[b]{0.45\textwidth}
			\includegraphics[width=1\textwidth]{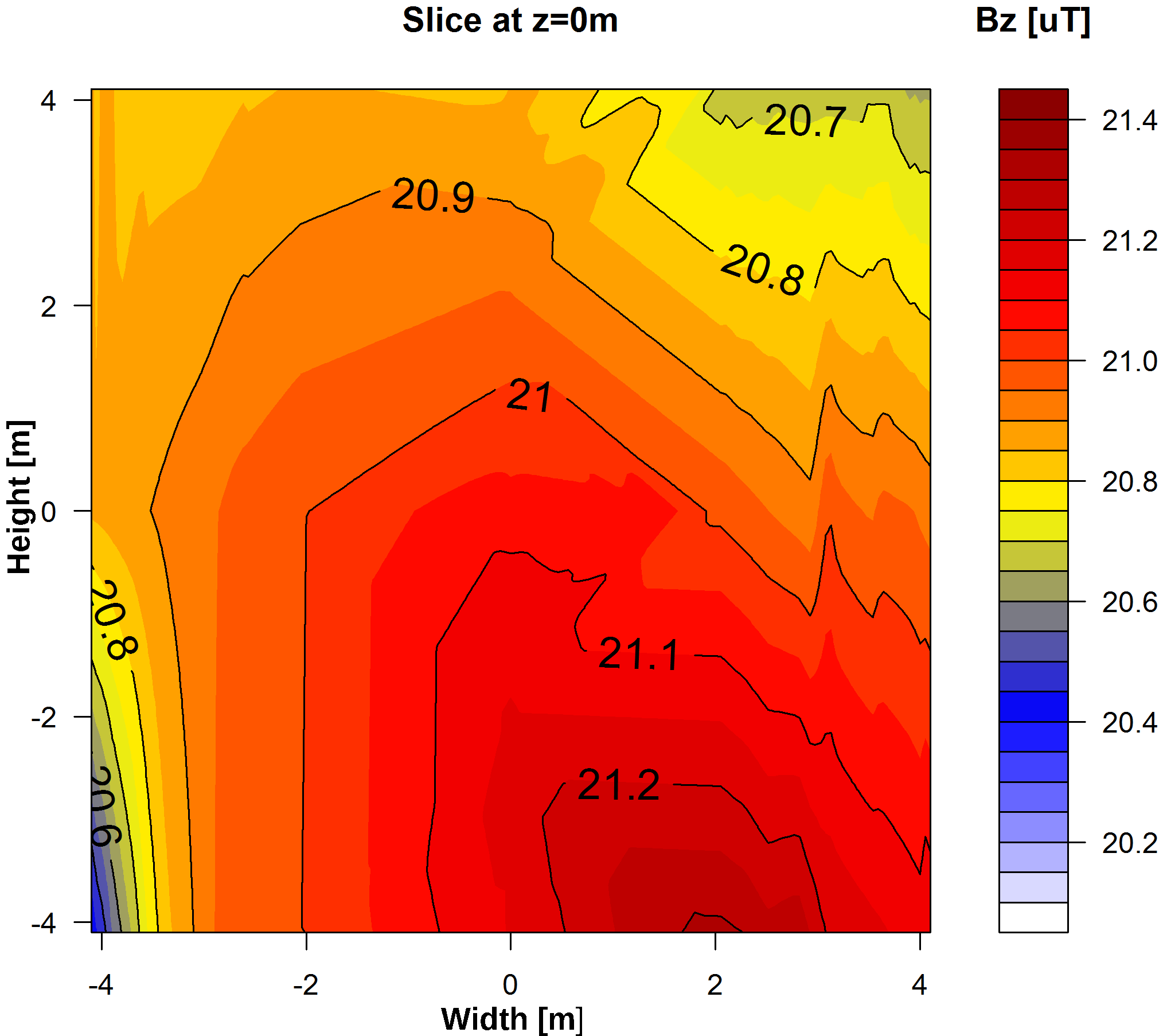}
			\caption{}
		\end{subfigure}
		\caption{Interpolation results of $B_z$ in the analysis plane (center of the MS) for (a) detector magnet at 2.5\,T, pinch magnet at 4.2\,T and PS at 3.1\,T, all other magnets off (b) all magnets off, only earth magnetic field.}
		\label{fig:MidSlice}
	\end{figure}

	The availability of the measured values in KATRIN coordinates 
	allows a direct comparison with simulated values. On the other hand one can use interpolation methods to derive magnetic field values inside the MS volume. The magnetic field in the analyzing plane as a result of a bi-linear interpolation on an irregular grid  performed in \cite{Let2018} is shown in \autoref{fig:MidSlice}. This method covers 86\% of the total analyzing plane area.
	
	\section{The vertical magnetic field measuring system}
\begin{figure}[h]
		\centering
		\includegraphics[width= .4\textwidth]{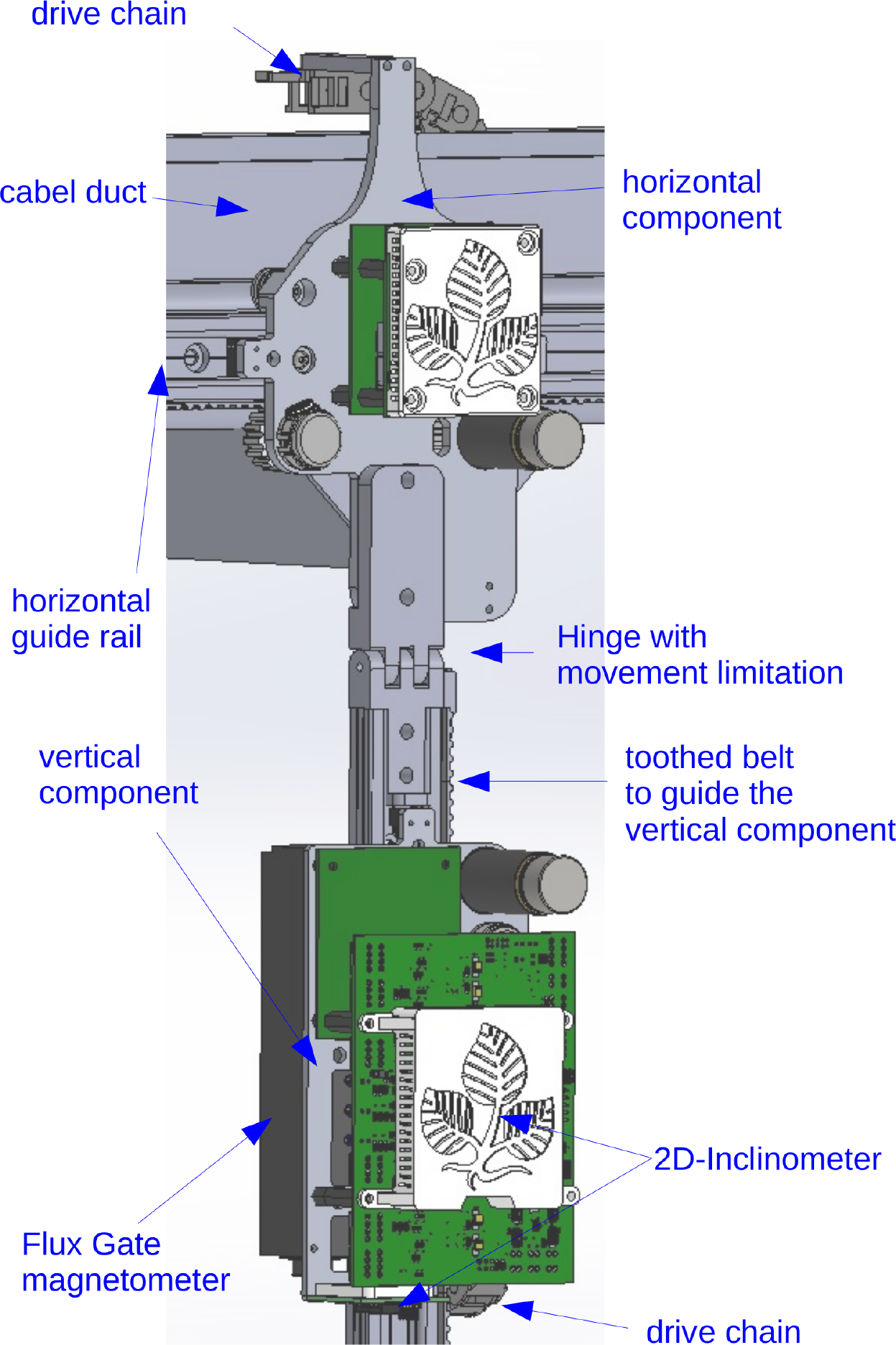}
		\caption{Schematic structure of the VMMS system in a CAD view.}
		\label{fig:VMMSmech}
	\end{figure}

\begin{figure}[h!]
		\centering
		\includegraphics[width=1\linewidth]{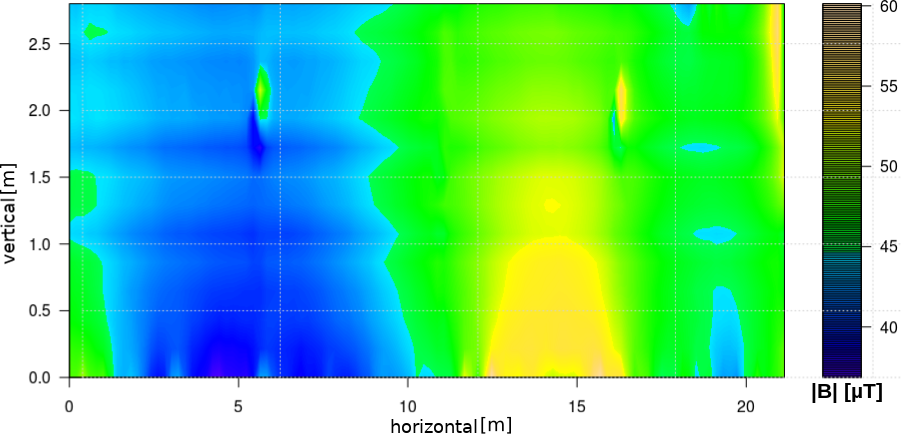}
		\caption{Interpolation result of the earth magnetic field measurement $|\vec{B}|$  in $\mu$T using 1484 sampling positions of the EC VMMS.}
		\label{fig:MagVInterpol_Earth}
	\end{figure}

	Magnetic field investigations in the immediate environment of the MS revealed  both remanent and induced magnetization effects of the hall walls which have a direct influence on the magnetic field in the analysis plane (see \cite{Erh2016}). For this reason, a vertical magnetic measuring system (VMMS)  covering vertical planes parallel to hall walls has been developed. Mechanically, the VMMS is inspired by the technology of the MobSU and is based on a movable construction of linear rails which are attached to the hall pillars. 
	\begin{figure}[h]
		\centering
		\includegraphics[width= .55\textwidth]{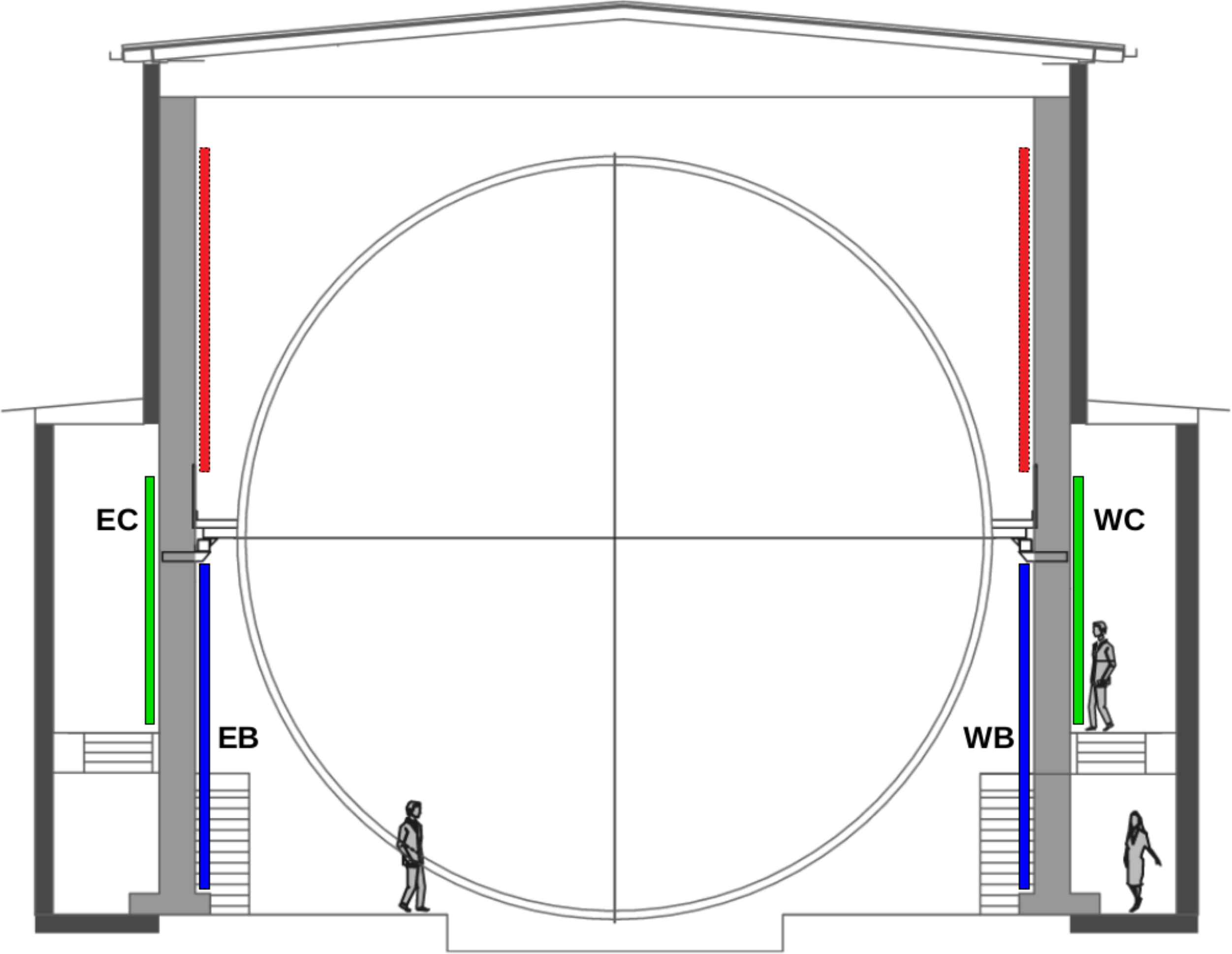}
		\caption{Schematic representation of the VMMS position within the KATRIN 
			MS hall. Individual layers are color-coded. View from the north. Individual systems are marked with EC (east center), EB (east bottom), WC (west center) and WB (west bottom).}
		\label{fig:VMMSHall}
	\end{figure}
	In terms of measuring accuracy and positioning precision, all requirements are met equivalent to the system of the mobile sensor units. The aim is to completely cover the wall surface in the area of the cylindrical MS vessel at three height levels and to measure the magnetic field with a mesh width of 20\,cm x 20\,cm. At the current stage four VMMS at two height levels are installed and commissioned. \autoref{fig:VMMSHall} shows the position of the individual systems. The construction of the upper system (in red) is currently in the concept phase and will be finished in the near future.\\ 
	\autoref{fig:VMMSmech} shows the schematic structure of such a VMMS system in a CAD view. The two movable components (horizontal and vertical), the movement limited hinge to prevent inadmissibly strong pendulum movement during the movement, as well as the drive chains connecting the subsystems and the cable duct can be seen. 
	The movement sequence is as follows: the vertical component starts in the lowest position and moves upwards to the next sampling position at a distance of 20\,cm. After the measurement has been carried out at this point, it moves to the next sampling position. As soon as the end of the vertical linear rail has been reached\footnote{this is indicated by a limit switch installed on the components}, the horizontal component is moved to the next position and the vertical system returns back to the initial point. After reaching the overall target position, the above procedure is repeated. In this way, a grid of magnetic field measuring points is built up until the end of the horizontal linear rail is reached. \autoref{tab:VKenngr} shows a listing of the most important parameters of the VMMS with regard to track length, total measuring time and total number of measuring positions. Using the already mentioned method of bilinear interpolation on irregular grids, a more detailed investigation of the influence of magnetization effects on hall walls can be carried out. Using the EC component of VMMS as an example, the interpolation results of the magnetic field magnitude $|\vec{B}|$ are shown in the \autoref{fig:MagVInterpol_Earth}.
	\begin{table}[h]
		\centering
		\begin{tabular}{|l|c|c|c|c|c|}
			\hline 
			& \textbf{EB} & \textbf{EC} & \textbf{WB} & \textbf{WC}  \\ 
			\hline 
			\hline
			vertical track length [m]   & 4.40 & 2.60 & 4.60 & 2.60 \\ 
			\hline 
			horizontal track length [m] & 20.20 & 21.00 & 20.80 & 21.00 \\  
			\hline 
			measuring positions &  2346 & 1484 & 2520 & 1484 \\ 
			\hline
			measuring time [h] & 5.68 & 3.85 & 6.16 & 3.85 \\ 
			\hline
		\end{tabular} 
		\caption{Significant parameters in relation to the track length, number of measuring positions and total measuring time of individual VMMS systems.}
		\label{tab:VKenngr}
	\end{table}

	\section{Summary and Outlook}
	In order to inspect the magnetic field inside the KATRIN main spectrometer during normal operation, two high-resolution magnetic sensor systems  have been developed and combined to form a fine-meshed magnetic sensor network. Depending on the mesh size, several thousand magnetic field samples can be taken close to the surface of the MS (RMMS) and on vertical planes left and right of the main spectrometer (VMMS). The electromechanical features of the mobile sensor units have been investigated and the serviceability of the systems has been demonstrated. 
	It has also been shown that the magnetic field at the analyzing plane $B_A$, formed by the MS spectrometer magnets inside the main spectrometer, can be derived from the RMMS magnetic field samples by interpolation. However, with more refined interpolation strategies and methods like the Laplace method \cite{Osi2012a} the stability of the result should be investigated. Moreover, as the error connected with interpolation numerics generally varies with $1/N$, with $N$ being the number of sampling points, the installation of 2 more units on LFCS 7 and LFCS 8 adjacent to the analyzing plane is advisable.\\ Furthermore, model based simulations of the magnetic field  can be checked and improved by magnetic field samples at the MS site. Especially the magnetic dipoles method \cite{Rei2014} needs several thousand B-field samples from RMMS and VMMS. Using this method the magnetization of model
	dipoles (several hundreds) can be determined and represented in simulations.

	
	\acknowledgments
	The authors wish to express gratitude to the group for Experimental Techniques of the Institute for Nuclear Physics (IKP) at KIT  and for highly efficient and competent support. Furthermore, we wish to thank Prof. Dr. E. W. Otten, Mainz University for helpful discussions and support. In addition, we like to thank the University of Applied Sciences, Fulda and the Fachbereich Elektrotechnik und Informationstechnik for their support during the entire process.
	
	This work has been funded by the German Ministry for Education and Research under the Project codes 05A14REA, 05A11REA, 05A08RE1
	
	%

	

\end{document}